\documentclass{article}
\title{Stochastic aggregation of point particles revisited.}
\author{Oleg V. Zaboronski\\
Department of Mathematics,
University of Warwick. \\ Coventry, CV4 7AL, UK; olegz$@$maths.warwick.ac.uk.}
\begin{document}
\maketitle
\newcommand{\thv}{{\bf \theta }}
\newcommand{\xv}{{\bf x}}
\newcommand{\nv}{{\bf n}}
\newcommand{\ek}{{\bf e}_{k}}
\newcommand{\ad}{a^{\dagger}}
\newcommand{\vac}{\mid 0 \rangle}
\newcommand{\nket}{\mid \nv \rangle}
\newcommand{\adth}{a_{\thv}^{\dagger}}
\newcommand{\ath}{a_{\thv}}
\newcommand{\svt}{\mid \Psi (t) \rangle}
\newcommand{\lvac}{\lange 0 \mid}
\newcommand{\phis}{\phi ^{*}}

\begin{abstract}
In the present letter we 
employ the method of the dynamical renormalisation group to
compute the average mass distribution
of aggregating point particles in $2$ dimensions in the regime when
the effects of local mass distribution fluctuations are essential.

\end{abstract}

The statistical properties of systems of aggregating
diffusive
particles have been studied using methods of kinetic theory
since the pioneering work of Smoluchowski, \cite{Smol}.
The kinetic equation derived in this work almost a hundred
years ago has proved extremely useful in describing
properties of various aggregation systems, see \cite{Ernst}
for a review. However, as it became clear in the $80$'s,
the Smoluchowski equation provides a sort of mean field
theory description of the process of aggregation and is
therefore valid if only the fluctuations of particle density
around mean values are small. 

Both theoretical analysis within the mean field theory
\cite{Dongen}
and numerical experiments \cite{Redner} discovered the 
existence of the $critical$ dimension $d_{c}$
of the ambient space
which marks the transition from non-fluctuating
large time statistics of a given model of aggregation
to a fluctuation-dominated one.
This means
in particular, that the scaling
properties of the large time asymptotics of average
mass distribution in $d>d_{c}$ are well described by an 
appropriate Smoluchowski equation,
whereas in $d<d_{c}$ the mean field theory answer for the large
time asymptotics of the average mass 
distribution becomes incorrect.

By analogy with equilibrium critical phenomena 
one should expect that mean field theory answers receive
logarithmic corrections in $d=d_{c}$. These have indeed been 
observed in numerical experiments \cite{Redner} and explained
heuristically \cite{Krap}.

However, the theory of stochastic aggregation
which would enable one to account for fluctuation effects 
systematically, in particular to derive the above mentioned
logarithmic corrections in $d=d_{c}$, is still absent.
 
The aim of the present letter is to build such a
theory 
for the simplest model
of stochastic
aggregation of diffusing point particles characterized by
constant diffusion and reaction rates.
This model was originally introduced in \cite{Redner}.
Its critical dimension was found to be equal to $2$.

We start the discussion by defining the model in terms
of the master equation. We then use Doi's formalism \cite{Doi} 
to derive the functional
integral representation for the average mass distribution.
We then establish the connection
between the resulting effective
field theory and the well understood effective field theory
of $A+A \rightarrow A$ model \cite{Peliti}. 
The connection is used to compute
the average density in $2-\epsilon$ dimensions. 
This allows an easy derivation
of logarithmic corrections to the mean field theory answers in $d=2$,
the determination of
the exact
scaling properties of the answer in $d=1$ and a discussion of
the mean field theory applicability in $d>2$.

Consider a set of point particles on a $d$-dimensional
lattice $Z^{d}$ each performing a random walk characterized
by a rate $D$. 

Suppose that each particle carries an integer
charge $q$. Any two particles 
with charges $q_{1}$ and $q_{2}$ positioned at a given lattice site
can coalesce with probability $\lambda$ per second to
create a particle with a charge $q_{1}+q_{2}$ in the same position. 

It is convenient to think of our particle system as living
on a $(1+d)$-dimensional lattice $\Gamma$
with $d$ space coordinates
and one charge coordinate.
We assume that initially the occupation numbers of particles
of a given charge at different lattice sites are independent
identically distributed Poisson random variables. We also assume
that at $t=0$ there was no particles with negative charges.
Then there will be 
no such particles at any later time and
our problem becomes identical to the problem of $mass$ aggregation.

We wish to study the
time evolution of
the average 
number of particles at a site of $\Gamma$ which
corresponds,
after taking the continuum limit, to the
mass distribution of particles.

A microstate of our particle system is defined by specifying
the number of particles (the occupation number)
at each site of $\Gamma$.
The evolution of the probability measure $\bf P$
on the space  of microstates is governed
by the so called Master Equation. 
Our further considerations are based on the following two facts
about the Master Equation:\\
\\
(i) The Master Equation is linear;\\
(ii) The Master Equation is the first order 
differential equation w. r. t. time.\\
\\
Thus following the already standard construction
of \cite{Doi} 
one can construct a path integral representation
of the solution to the Master Equation and, consequently,
of any correlation function. 
Leaving details of the derivation to 
\cite{Oleg} we present here the final path integral expression
of the average mass distribution in
the continuum limit and study it.

Let $\Delta x$ be the lattice spacing in the coordinate space,
$\Delta q$ - in the mass space. The continuum limit is taken 
according to the following rules:
\begin{eqnarray}
\frac{\lambda}{2} (\Delta x)^d \rightarrow  \lambda _{0},~
D (\Delta x)^2 \rightarrow D.
\end{eqnarray}
Then the average density is
\begin{eqnarray}
n(q,t) = \int D\phi (\xv, q, t) D\phis (\xv, q, t) 
\phi (\xv, q, t) e^{-S[\phi, \phis]+ 
\int d^dx \int dq n_{0} (q) \phis(\xv ,q)},
\label{ad}
\end{eqnarray}
where 
\begin{eqnarray}
S= \int_{0}^{t} dt \bigg\{
\int d^{d} x dq  \phis (\xv , q, t) \partial _{t} 
\phi (\xv , q, t) 
+H [ \phi, \phis ] \bigg\} 
\label{ea}
\end{eqnarray}
is an effective action,
\begin{eqnarray}
H = D \int d^{d} x \int dq \nabla \phis (\xv,q) 
\cdot \nabla \phi (\xv,q)- 
\lambda _{0} \int d^d x \int \int dq dq' 
\nonumber \\
\bigg(
\phis(\xv , q+q') -2\phis (\xv, q)- \phis(\xv, q) \phis (\xv, q')
\bigg) \phi (\xv, q) \phi (\xv, q')
\label{eh}
\end{eqnarray}
is an effective Hamiltonian;
$n_{0} (q)$ is the initial average mass distribution,
\begin{eqnarray}
n (q,t) = \frac{4 n_{0}^3}{M^2} q e^{-\frac{2q n_{0}}{M}}
\label{id}
\end{eqnarray}
if $q>0$ and $0$ otherwise. Here $n_{0}$ is the total
initial
particle density, $M$- the total mass density. 

The  non-hermiticity of the effective
Hamiltonian
reflects the irreversibility of the
process of aggregation.
As in case of the equilibrium
statistical mechanics,
our problem has been mapped to an
effective Euclidean field theory 
in $d+1$ dimensions. But, unlike the
equilibrium case, the parameter $t$  has the
meaning of time, not
inverse temperature.

A simple scaling analysis of (\ref{ea}) yields the applicability
conditions of the mean field theory approximation in $d<2$:
The rescaling

\begin{eqnarray}
t \rightarrow t \tau, ~ \xv \rightarrow \sqrt{Dt} {\bf \xi},~
\phis \rightarrow \phis ,~ \phi \rightarrow Z \phi, ~
q \rightarrow W q
\label{resc}
\end{eqnarray}
with $ZW=\frac{1}{\lambda _{0} t}$, 
transforms the functional integral measure as
follows:
\begin{eqnarray}
e^{-S[\phi, \phis, \lambda_{0}, t, D]} =
e^{-\frac{1}{g(t)} S[\phi, \phi ^{*}  , 1, 1, 1]} ,
\label{im}
\end{eqnarray} 
where 
\begin{eqnarray}
g(t)= \frac{\lambda_{0} t}{(Dt)^{d/2}}
\label{dc}
\end{eqnarray}
is the dimensionless coupling.

We see that the
mean field theory (or, equivalently, the weak coupling ($g(t)<<1$)
approximation ) is always violated  
in the limit 
$t \rightarrow
\infty $
if $d<2$. Dimension $2$ is critical in a sense that
the dimensionless coupling $g \mid _{d=2}$ doesn't depend on time. 

In the weak coupling limit
the  average mass distribution
$n(q,t)$ solves the
Euler-Lagrange equation describing the extrema of
the effective action (\ref{ea}):

\begin{eqnarray}
\partial _{t} n(q,t) =\frac{\lambda_{0}}{2} 
\int_{0}^{q}dq' n(q-q',t)n(q',t) 
-\lambda_{0} \int_{0}^{\infty} dq' n(q,t) n(q',t),
\label{sm}
\end{eqnarray}
in which we recognize the Smoluchowski equation.

Now we can explain physical
meaning of the 
condition $g(t)<<1$ of applicability of the
Smoluchowski equation.
Integrating (\ref{sm}) w. r. t. $q$ one finds
that the $total$ density  $n(t) = \int dq n(q,t)$
obeys the 
standard rate equation of the theory of $A+A \rightarrow A$
reaction:
\begin{eqnarray}
\partial _{t} n = -\frac{\lambda_{0}}{2} n^2,~ n(0)=n_{0}.
\label{re}
\end{eqnarray}
The asymptotic solution of this equation for
$t>>n_{0}^{-1} \lambda_{0} ^{-1}$ is  $n(t) \sim \frac{1}{\lambda_{0} t}$.
Note that $n(t)^{-1/d}$ is the average distance between particles
which determines the correlation length and $\sqrt{Dt}$ is the typical
size of spatial fluctuations of local density. Thus the condition
$g(t)<<1$ is equivalent to
\begin{eqnarray}
\frac{l_{corr}}{l_{diff}} <<1.
\label{gl}
\end{eqnarray}
The conclusion is
that if the correlation length is much smaller than
the scale of fluctuations, the spatial inhomogeneity
of local density is irrelevant and the
mean field theory is applicable.
Condition (\ref{gl}) should be compared
to the Ginsburg-Landau
criterium of applicability of mean field theory
approximation
to
the Ising model, see \cite{Cardy} for a review.
Note that in $d<2$
the correlation length measured in units of a typical
fluctuation size diverges as $t \rightarrow \infty$.
Spatial correlations become important and mean field
theory breaks down in complete analogy with 
the break up of mean field theory approximation of
equilibrium statistical systems near the critical point.

Analyzing the applicability
of the mean field theory we concentrated
on parameters $\lambda_{0}$ and
$t$ only, keeping the rest of the parameters 
of the theory fixed. 
As a result, (\ref{gl}) gives only
a necessary condition of the applicability of Smoluchowski 
equation.  In particular, we will see below 
that the Smoluchowski theory
of aggregation can produce incorrect results for very small 
and very large $q$'s even if (\ref{gl})
is satisfied.

Smoluchowski equation (\ref{sm})  with the initial
condition (\ref{id}) is easy to solve using Fourier transform
in $q$ variable with the following result for the
average mass distribution:
\begin{eqnarray}
n(q,t)=\frac{2 n_{0}^2}{M F^2 (t)} \frac{sinh(\frac{2 q n_{0}}{M} 
\sqrt{1-F^{-1}(t)})}{\sqrt{1-F^{-1}(t)}} e^{-\frac{ 2 q n_{0}}{M}},
\label{mfs}
\end{eqnarray}
where $F(t)=1+n_{0} t \lambda _{0}$.
Note that $n_{0} \rightarrow \infty$ limit of (\ref{mfs})
exists and corresponds to the unique self-similar solution
of (\ref{sm}):
\begin{eqnarray}
lim_{n_{0} \rightarrow \infty} n(q,t) = 
\frac{1}{M \lambda _{0}^2 t^2} e^{-\frac{q}{ M \lambda _{0} t}}
\label{us}
\end{eqnarray}

To compute the average mass distribution taking into account
the fluctuation effects we
must analyze the full effective field theory (\ref{ea}).
This turns out to be a relatively simple task due to the
connection between (\ref{ea}) and the effective field
theory of $A+A\rightarrow A$ which we will now describe.

It is well known that the evolution of total density
of particles
in constant kernel stochastic aggregation
is governed by $A+A\rightarrow A$ process. Here we
establish   
a more general result: {\em the average mass distribution
in constant kernel aggregation is given by
the Fourier transform of the (analytical continuation of)
average density of particles in $A+A\rightarrow A$ process
with respect to the initial average density}. 

To verify the statement we note by
analogy with \cite{Cardy1} that
the last term  in the effective Hamiltonian (\ref{eh}) is an 
$x$-integral of the complete square of the field $W= \int dq \phis \phi$.
The appropriate term in the 
functional integral measure in (\ref{ad}) can thus be rewritten
using the Hubbard-Stratonovich
transformation:
\begin{eqnarray}
e^{- \lambda \int dx dt W^2} = \int D\xi (x,t)
e^{-\frac{1}{2} \int dx dt (\xi ^2 \pm i \sqrt{\lambda} W\xi )}.
\label{hs}
\end{eqnarray}

Substituting (\ref{hs}) into (\ref{ad}) we find that the exponent
of the new functional integration measure  
is $linear$ in $\phis$. 
The quadratic part of the new effective action is non-degenerate.
Thus the
$\phis$-field can be integrated out, at least perturbatively. Following
the inverse of Martin-Siggia-Rose (MSR) procedure \cite{MSR}
one finds that the computation of the average mass distribution is equivalent
to averaging the solution of the following stochastic integro-differential
equation with respect to noise:
\begin{eqnarray}
(\partial _{t}-D\Delta) \phi (\xv ,t,q) =\frac{\lambda_{0}}{2}
\int_{0}^{q}dq' \phi (\xv , t,q-q') \phi (\xv ,t, q') \nonumber \\
-\lambda_{0} \int_{0}^{\infty} dq' \phi (\xv , t, q) \phi (\xv , t,q')
-2i \sqrt{\lambda_{0}}\cdot \xi (\xv, t) \phi (\xv, t, q),
\label{ssm}
\end{eqnarray}
where the stochastic potential $\xi (\xv ,t)$ is Gaussian and
"white", $\langle \xi (\xv ,t) \xi (\xv ',t') \rangle = 
\delta (\xv -\xv ')\delta(t-t')$. Imaginary unit multiplying the noise term
of (\ref{ssm}) is due to the fact that aggregating particles are
$anticorrelated$. Note also that 
that even though $ n (q,t)  =\langle \phi (q,t) \rangle$,
$\phi$ and  the local mass distribution
are $different$ as random variables. For instance,
all higher moments of $\phi$-distribution 
are different from higher moments of $n$-distributon. See \cite{Cardy1}
for more details.

Rescaling (\ref{resc}) 
maps (\ref{ssm}) to an equation with $g(t)$ entering only
as a multiplier of the noise term. Thus in the weak coupling
limit (\ref{ssm}) reduces to (\ref{sm}), as it should.

Equation (\ref{ssm}) simplifies upon taking a Fourier transform
with respect to the mass variable $q$:
\begin{eqnarray}
(\partial _{t} -D\Delta)\phi (\xv, t, k) 
\nonumber \\
=2\pi \lambda _{0}
\phi ^{2} (\xv , t, k) +4\pi \lambda _{0} \phi (\xv, t, k)
+2i \sqrt{\lambda _{0}} \xi (\xv ,t) \phi (\xv, t, k).
\label{ftsm}
\end{eqnarray}

Hence the random field $\phi (\xv, t, 0)$ which corresponds
to the total density does indeed
satisfy the closed stochastic partial differential
equation (SPDE) arising in
the theory of $A+A \rightarrow A$ reaction, \cite{Cardy1}.
In general, the field $\phi (\xv,t,k)$ is coupled only to itself and
$\phi (\xv ,t,0)$. Introducing new random variables
$A(\xv ,t) = \phi (\xv  , t, 0)$ and 
$B(\xv,t) = \phi (\xv ,t,0) - \phi (\xv,t,k)$ we find that
they satisfy the following SPDE's:
\begin{eqnarray}
(\partial _{t} -D\Delta)A (\xv, t) +2\pi \lambda _{0}
A ^{2} (\xv , t)
-2i \sqrt{\lambda _{0}} \xi (\xv ,t) A (\xv, t)=A_{0} \delta (t)
\label{a}\\
(\partial _{t} -D\Delta) B (\xv, t) -2\pi \lambda _{0}
B ^{2} (\xv , t) 
-2i \sqrt{\lambda _{0}} \xi (\xv ,t) B (\xv, t)=B_{0}\delta (t),
\label{b}
\end{eqnarray}
where $A_{0}=\frac{n_{0}}{2\pi},~ B_{0}=\frac{n_{0}}{2\pi}
(1-(1+\frac{ikM}{2n_{0}})^{-2})$ are the initial values
of random variables $A$ and $B$ correspondingly. 

But if $q>0$, $n(q,t) =-\int dk e^{ikq} \langle B (\xv,t) \rangle$,
where $B$ is the solution of SPDE (\ref{b}) of the $A+A \rightarrow A$
theory with the initial density parametrised by $k$. In the universal
limit $n_{0} \rightarrow \infty$ the initial "density" is 
\begin{eqnarray}
B_{0} =ikM,
\label{b0}
\end{eqnarray}
thus 
confirming the statement about the connection
between the constant kernel aggregation and the theory of $A+A\rightarrow A$
made earlier. As $n_{0}$ flows to infinity under the renormalisation
group transformation \cite{Peliti}, we can set it equal to infinity
from the very beginning without affecting the answers for $t >> \frac{1}{\lambda_{0} n_{0}}$.
Hence everywhere below we assume that $B_{0}$ is given by (\ref{b0}).

The random field $A$ turns out to be irrelevant for the computation 
of the average mass distribution. Yet it is not completely decoupled
from the $B$-field due to the common random potentials 
of SPDEs (\ref{a}) and (\ref{b}). This remark is important for the computation
of correlation functions, \cite{Oleg}.

The Fourier transform of the average mass distribution
$n(k,t)$ is most conveniently computed within the field-theoretic
formalism. Thus we apply the MSR procedure to  (\ref{b})
to arrive at the following representation of $n(k,t)$:
\begin{eqnarray}
n(k,t) =n(t) -  \int DB B e^{-S[B]+\int d^{d} x B^{*} B_{0}},
\label{ad1}
\end{eqnarray}
where $n(t)$ is the average density of particles and 
\begin{eqnarray}
S[B] = \int_{0}^{t} dt \int d^d x \{
B^{*} (\partial _{t} -D\Delta) B 
+2\pi\lambda _{0} B^{*} B^2 + \lambda _{0} (B^{*} B)^2 \}
\label{aa}
\end{eqnarray}
is an effective action of the $A+A \rightarrow A$ theory, \cite{Peliti}.

Now we shall obtain a representation of 
$n(k,t)$ as  a perturbative series in $g(t)$, identify
the terms in the series giving the leading order 
contributions in the large $t$-limit
and sum these contributions exactly using the formalism 
of renormalization group.
In our analysis
we are going to use fairly standard methods of the statistical field
theory, see e. g. \cite{Cardy}, \cite{Drouffe} for a review. 

The 
Fourier transform of the average mass distribution can
be presented
graphically as a sum of all Feynmann diagrams of the theory 
(\ref{aa}) with
a single outgoing line. These can be classified
by a number of loops. To regularise the divergencies
due to integration along loops we use the dimensional regularization,
i. e. compute Feynman integrals associated
with each diagram in dimension $d=2-\epsilon$, where
$\epsilon$ is a small parameter. A simple combinatorial and
dimensional analysis shows that a graph with $n$ loops and
$m$ ingoing density lines yields a contribution to the 
average density of the form   
\begin{eqnarray*}
C_{n,m} (\epsilon) \lambda^{n-1} (Dt)^{-d/2-(n-1)(d-2)/2}
(\lambda t B_{0})^{m},
\end{eqnarray*}
where $C_{n,m} (\epsilon)$ is a constant.
Therefore,
\begin{eqnarray}
n(k,t) =n(t)- \frac{1}{\lambda_{0} t}
\sum _{n=0}^{\infty} g(t)^{n} f_{n} ( 
\epsilon, \kappa),
\label{ads}
\end{eqnarray}
where $\kappa =-i\lambda_{0} B_{0} \equiv \lambda_{0} t M k$ is
a {\em scaling variable}: For a fixed
$\kappa$ expansion (\ref{ads}) is indeed an expansion in powers of $g(t)$.
An expansion coefficient $f_{n} (\epsilon, \kappa)$ is given by
the sum of all Feynman diagrams containing precisely $n$-loops.
As $\kappa$ is not a small parameter of the theory, these sums have to
be computed exactly.
Direct evaluation of tree and 1-loop contributions gives:

\begin{eqnarray}
n( \kappa ,t) = n(t)-\frac{1}{\lambda_{0}t} 
\bigg(
\frac{i\kappa}{1+i\kappa}
 - \frac{2}{\pi \epsilon}
\left( \frac{1}{8\pi}
\right) ^{d/2} g(t) 
\frac{\kappa^2}{(1+i\kappa)^2} 
\nonumber \\
\times \int_{0}^{1} \frac{dx}{(1+i\kappa x)^2} x^{\epsilon /2} 
\left( 1+
\frac{4i\kappa x}{(2+ \epsilon)}  
-\frac{8 \kappa^2 x^2}{(2+\epsilon)(4+\epsilon)}  
\right)
+ O
\left( g(t)^2 
\right) 
\bigg)
\label{1lpad}
\end{eqnarray}

Note that the sum of all
$tree$ diagrams yields the solution of the
rate equation (\ref{re}) with the initial condition (\ref{b0}).
Hence the expansion (\ref{ads}) is the standard loop expansion around
a nontrivial classical solution. 

The inverse Fourier transform 
of (\ref{1lpad}) produces the $g(t)$-expansion of the average mass 
density around the solution (\ref{us}) of the Smoluchowski 
equation. The coefficients of this expansion depend on
the mass scaling variable  $Q=q/(\lambda_{0} t M)$. 

We see that if $g(t)<<1$, there are domains in $\kappa$
and $Q$ spaces
in which the mean field theory answer holds.
If however $d<2$ they shrink
with time
due to the the growth of the dimensionless coupling $g(t)$. 
Note that we do not discuss the situation in $d>2$, where our
considerations must be modified  to 
account for the strong cut-off
dependence of the answers.

The growth of $g(t)$ is in turn due to the recurrent
property of random walks in $d<2$: the probability that
two particles coalesce at some late
time $t$ is strongly modified
by the fact that they met  many
times in the past and yet didn't react, \cite{Doering}. In other words,
the bubble diagrams renormalising the coupling $\lambda_{0}$
are all relevant at large times, \cite{Peliti}. 
A simple dimensional argument shows that only 
the renormalisation of two particles
processes is relevant at large times in $d<2$. Thus 
one can extract the large
time asymptotics of the average mass distribution  from (\ref{ads})
by rearranging the $g(t)$-expansion in such a way that it
becomes an expansion w. r. t. a renormalised dimensional
coupling $g_{R} (t)$. This is precisely the task 
the renormalisation group method has been designed for.

The contributions corresponding to diagrams renormalising $\lambda_{0}$
form a geometric sequence and thus can be summed exactly. The
result is
\begin{eqnarray}
 \lambda_{R} (t) =\frac{\lambda_{0}}{1+ 
\frac{C_{d}}{\epsilon} \lambda_{0} t (Dt)^{-d/2}} ,
\label{lr}
\end{eqnarray}
where $C_{d}$ is a positive constant.
Suppose that the average mass distribution at $t=t_{0}$
is known. Then we can in principle compute it for any later time $t$
by solving the field theory (\ref{aa}), $but$ with $\lambda_{0}$
and $n_{0} (k)$
replaced with $\lambda_{R} (t_{0})$ and
$n(k,t_{0})$ correspondingly. 
The result for $t>t_{0}$
will not
depend on $t_{0}$, as the evolution operators $U(t',t'')$ corresponding
to (\ref{aa}) form a semi-group: $U(0,t_{0}) \cdot U(t_{0},t)=U(0,t)$.
With the aid of dimensional analysis condition $\partial_{t_{0}} n(k,t)=0$
can be written  in the form of 
Callan-Symancyk (renormalization group) equation:
\begin{eqnarray}
\bigg( t\frac{\partial}{\partial t}+\frac{1}{2} 
\beta(g_{R}) \frac{\partial}{\partial g_{R}}
-\frac{d}{2} k\frac{\partial}{\partial k} +
\frac{d}{2} \bigg) n\big( k,t, g_{R}(t_{0}) \big)=0, ~t>t_{0}.
\label{cs}
\end{eqnarray}
Here 
\begin{eqnarray}
\beta (g(t)) \equiv -t \frac{\partial \lambda _{R} (t)}{\partial t}
=C_{d} g(t)^2 - \epsilon g(t) 
\label{bf}
\end{eqnarray}
is an exact $\beta$-function of the field theory (\ref{aa}).  
The solution of Calan-Symancyk equation is 
\begin{eqnarray}
n(k,t) =\bigg( \frac{t_{0}}{t} \bigg) ^{d/2} 
n \bigg( \bigg( \frac{t_{0}}{t} \bigg) ^{d/2} k, t_{0} , g_{R} (t,t_{0}) \bigg),
\label{gs}
\end{eqnarray} 
where $g(t,t_{0})$ is the {\em running coupling constant}:
\[
g_{R}(t,t_{0})= \left\{ \begin{array}{l@{\quad:\quad}lrr}
\frac{\epsilon}{C_{d}} \big( 1-(1-\frac{\epsilon}{C_{d} g_{R}(t_{0})}) 
(\frac{t_{0}}{t})^{\epsilon/2} \big) ^{-1} & d<2 & \hspace*{3.8cm}&(27.1)\\
g_{R}(t_{0}) \big( 1+
\frac{g_{R}(t_{0})}{4\pi}ln(\frac{t}{t_{0}}) \big) ^{-1} & d=2
& &(27.2)
\end{array} \right. \]

Expressions (\ref{1lpad}) and (\ref{gs}) 
can be used in the analysis of
the large time asymptotics of the average mass distribution
in the following way:

\noindent
If $d<2$, $\lim_{t\rightarrow \infty} g_{R} (t,t_{0}) =\frac{\epsilon}{k_{q}}$, which
is a non-trivial fixed point of the $\beta$-function (\ref{bf}).
Thus the large time asymptotics of the average mass distribution 
in $d=1$ has according
to (\ref{gs}) the following self-similar form:
\begin{eqnarray}
n(q,t) = \frac{M}{Dt} \Phi \bigg( \frac{q}{M\sqrt{Dt}} \bigg),
\label{d1}
\end{eqnarray}
where $\Phi$ is some universal scaling function. 
Hence our analysis reproduces
the exact scaling properties of the average mass distribution in
one dimension
known from the exact solution, \cite{Spouge}. The substitution
of (\ref{1lpad}) into (\ref{gs}) gives a
representation of $n(q,t)$ in the form of
$\epsilon$-series. 
Unfortunately, the $\epsilon$-expansion
doesn't reveal any information about the scaling function $\Phi$ when
$\epsilon =1~(d=1)$.

If however $d=2$, then 
$g_{R} (t,t_{0}) \sim 4\pi / ln (t/t_{0})<<1$ for $t>>t_{0}$.
Hence it is possible to evaluate
the r. h. s. of (\ref{gs}) for
large times using the weak coupling expansion (\ref{1lpad})! 
Technically, this possibility is realized as follows:
We express $g(t_{0})$ in the series (\ref{1lpad}) 
through $g_{R} (t_{0})$ using (\ref{lr}) for $\epsilon >0$
and substitute
it into (\ref{gs}). 
The resulting expression is non-singular at $\epsilon =0$, as the
renormalization of the coupling constant $\lambda_{0}$ takes care
of all perturbative divergencies of the field theory (\ref{aa}).
Thus we can set $\epsilon =0$ and after
taking the inverse Fourier transform obtain the following answer
for the average mass distribution in $d=2$:
\begin{eqnarray}
n_{R}=\frac{1}{M} (\frac{1}{g_{R}(t,t_{0})Dt})^2e^{-Q_{R}} (1+\nonumber 
\frac{g_{R} (t,t_{0})}{8\pi}
((2C-5)(2-Q_{R})+\nonumber\\
(2(1-Q_{R})(ln(Q_{R}+C)+2Q_{R}+1-\frac{1}{2}
(Q_{R}-2)^2))+O(g_{R}(t,t_{0})^2)),
\label{rad}
\end{eqnarray}
where $C$ is the Euler constant and $Q_{R}=\frac{q}{M g_{R} (t,t_{0}) Dt}$
is the renormalized scaling variable. The cutoff time
$t_{0}$ can be related to the lattice spacing: $t_{0} = \Delta x^2/D$.
Even though individual terms in the expansion 
(\ref{rad}) depend on $t_{0}$, their sum doesn't.
A simple check shows that integrating (\ref{rad}) with respect
to $q$ gives the one-loop approximation of the total density
obtained in \cite{Peliti}, while the integral of $q\cdot n(q,t)$ gives
the total mass density $M$. The expression
for the average mass distribution
can be simplified
even further if we use a more natural scaling variable
$Q=q\cdot n(t)/M$, where $n(t)$ is the exact total density.
Then the term $\frac{g_{R} (t,t_{0})}{8\pi}
(2C-5)(2-Q_{R})$ disappears from the r. h. s. of (\ref{rad}) and
we are left with the non-trivial
$form$ corrections to the scaling function $e^{-x}$.

Thus, we conclude that if
$Q_{R}>>t_{0}/t$ and $Q_{R} <<(ln(t/t_{0})^{1/2}$,
\begin{eqnarray}
n_{R} \approx \frac{1}{M} 
\bigg( \frac{ln(t/t_{0})}{4\pi Dt} \bigg) ^2
exp \bigg\{- \frac{qln\big( t/t_{0}\big) }{4\pi DtM} \bigg\},
\label{rmf}
\end{eqnarray}
thus establishing the result obtained numerically in \cite{Redner}
and conjectured on the basis of heuristic arguments in \cite{Krap}.
Note that (\ref{rmf}) holds at large times irrespectively of
the value of the coupling $\lambda_{0}/D$. But
if the latter is large, the mean field theory result (\ref{1lpad})
isn't valid at any time at all.

We emphasize the non-uniformity of the corrections
to the renormalized mean field answer (\ref{rmf}) with respect to 
the scaling variable $Q_{R}$. This phenomenon occurs in higher
dimensions as well. A computation shows that in $d<4$, the large-$Q$
asymptotics of the one-loop correction to the mean field result
behaves like $g(t) Q^2e^{-Q}$. Thus the mean field theory
can be violated even above the critical dimension if $Q$ is large enough.

The main reason for the violation of (\ref{rmf}) for large $q$'s
is the existence of another length scale in our problem,
$l_{M}=(q/M)^{1/d}$ (so far our analysis of the applicability
of the mean field theory used only the diffusion scale 
$l_{D}=\sqrt{Dt}$). Corrections to the mean field theory which
depend on $l_{M}$ are due to the fluctuations in the {\em flux
of masses} in the mass space due
to aggregation, see \cite{Oleg} for more details.

In conclusion I would like to stress that the applicability
of the statistical field theory methods to the stochastic aggregation
is not necessarily
limited to the simple model at hand. In particular, it would
be interesting to apply the machinery of renormalization
group to the model of stochastic aggregation with a multiplicative
kernel $K(q,q')=qq'$, 
for which the mean field theory predicts a finite
time gelation transition, \cite{Dongen1}.
Another potentially interesting application is to
a model of stochastic aggregation with fragmentation
which admits a non-equilibrium phase transition
between a trivial and a non-trivial steady states, \cite{Sup} \\
{\bf Acknowledgements.} I would like to thank John Cardy for 
useful discussions. The hospitality of the Department of 
Theoretical Physics of the University of Oxford 
where the present work has been completed
is greatly appreciated.
This research was partially supported by the grant No. 4603 from London
Mathematical Society.

\end{document}